\documentclass[12pt]{article}
\usepackage[pdftex]{graphicx}
\usepackage[english]{babel}
\usepackage{amssymb,epsfig}
\usepackage{amsmath}
\usepackage{slashed}
\usepackage[active]{srcltx}
\newcommand{\GeV}{\operatorname{GeV}}

\newcommand{\be}{\begin{equation}}
\newcommand{\ee}{\end{equation}}
\newcommand{\ba}{\begin{eqnarray}}
\newcommand{\ea}{\end{eqnarray}}

\begin{document}

\begin{flushright}
\end{flushright}

\bigskip

\begin{center}
\textbf{\LARGE  Form factors in $\bar{B}^0\to \pi^+\pi^0\ell\bar\nu_\ell$ 
from QCD light-cone sum rules}

\vspace*{1.6cm}

{\large Christian Hambrock and Alexander Khodjamirian}

\vspace*{0.4cm}

\textsl{%
Theoretische Physik 1,
Naturwissenschaftlich-Technische Fakult\"at,\\
Universit\"at Siegen, D-57068 Siegen, Germany}
\vspace*{0.8cm}

\textbf{Abstract}\\[10pt]
\parbox[t]{0.9\textwidth}{
The form factors of the semileptonic 
$B\to \pi\pi\ell\bar\nu$ decay
are calculated from QCD light-cone sum rules
with  the distribution  amplitudes of dipion states.
This method is valid in the kinematical region, 
where the hadronic dipion state has 
a small invariant mass and simultaneously a large recoil. 
The derivation of the sum rules is  
complicated by the presence of an additional variable
related to the angle between the two pions.
In particular, we realize that not all invariant amplitudes 
in the underlying correlation function can be used, some of them  
generating kinematical singularities in the dispersion relation.
The two sum rules that are free from these ambiguities 
are obtained in the leading twist-2 approximation,
predicting the $\bar{B}^0\to \pi^+\pi^0$ form factors
$F_{\perp}$ and $F_{\parallel}$ of the  
vector and axial $b\to u$ current, respectively.
We calculate these form factors 
at the momentum transfers $0<q^2\lesssim 12 $ GeV$^2$
and at the dipion mass close to the threshold $4m_\pi^2$.  
The sum rule results indicate that the contributions of 
the higher partial waves to the form factors are suppressed 
with respect to the lowest $P$-wave contribution and that the  
latter is not completely saturated by the $\rho$-meson term. 
}
\end{center}

\vspace*{1cm}

\newpage

\section{Introduction}
The current tendency in the studies of the flavour-changing decays
of heavy hadrons is to enlarge the set of exclusive processes  
used for the determination of the fundamental CKM parameters.
Probing different exclusive $b\to u$ processes may, in particular, 
help in the $|V_{ub}|$ determination.  The interval of 
this CKM parameter obtained from 
the measurements of the $B\to \pi\ell\bar\nu_\ell$ decay, combined with the 
$B\to \pi$ form factors from lattice QCD or 
from the  QCD light-cone sum rules (LCSR), deviates
from the results obtained  
in the inclusive $B\to X_u\ell\bar\nu_\ell$ decay studies (see, e.g., the review 
\cite{Bfactbook} and references therein). 

Alternative exclusive $b\to u$ processes are  being 
actively investigated, among them  the $B\to \pi\pi \ell\bar\nu_\ell$ decay,
where the $\rho$-meson contribution is prominent.
The semileptonic $B$-decay mode with the two-pion ({\em dipion}) final state is   
not only important for  
the $|V_{ub}|$ determination, but  also has  a  rich set of   
observables (see e.g., Ref.~\cite{FFKMvD}) which can be used for
nontrivial  tests of Standard Model. 
The $B\to \pi\pi \ell\bar\nu_\ell$ decay has already been measured, but mainly
its resonant, $B\to\rho\ell\bar\nu_\ell$ part (see e.g., the BaBar \cite{BaBarBrho} and Belle 
\cite{BelleBrho} collaborations data). 
Significantly more detailed data on the $B\to \pi\pi \ell\bar\nu_\ell$ 
observables are expected from the Belle-2 experiment in future.

The dynamics of the $B\to \pi\pi \ell\bar\nu_\ell$ decay
is governed by general $B\to 2\pi$ form factors, hence the calculation of these 
form factors is becoming the next big task for the practitioners of QCD-based 
methods. As  discussed in Ref.~\cite{FFKMvD} in detail, 
various non-lattice methods, from heavy-meson chiral perturbation theory 
to the soft-collinear effective theory are applicable, 
depending on the region of the Dalitz plot
formed by the invariant masses of the lepton pair and 
dipion.

In this paper, we use the method  of LCSRs  \cite{lcsr}
to calculate  the $B\to 2 \pi$ form factors relevant for 
the $\bar{B}^0\to \pi^+\pi^0 \ell^-\bar{\nu}_\ell$ decay. We
shall confine ourselves with the charged dipion (isovector) final state,  
and postpone the case of the neutral (isoscalar) state with related scalar resonances
for the future work. 
The approach we use is applicable in the region of small and intermediate 
lepton-pair masses, restricting simultaneously the dipion invariant 
mass by the $\lesssim 1$  GeV region, so that a large hadronic recoil takes place 
with two energetic and almost collinear pions  in the $B$-meson rest frame.  

The technique we use 
has many similarities  with the LCSRs obtained for $B\to \pi$ form factors, 
but employs a different and more complicated nonperturbative input: 
the light-cone 
distribution amplitudes (DAs) of the dipion 
state. These universal objects have been introduced 
in Refs.~\cite{2pionDA,Mueller:1998fv} to encode the hadronization
of the quark-pair in the   $\gamma\gamma^*\to 2 \pi$ process
at large momentum transfer.  The properties of dipion DAs were worked out in details in Ref.~ \cite{PolyakovNP,DGP}.
In a different context, two-meson wave functions in hard exclusive processes were 
discussed earlier in Ref.~\cite{Grozin}.

In this paper we aim at the following goals. First, we demonstrate how the method works, 
deriving the LCRSs for the two of the $B\to \pi\pi$  form factors in the 
leading twist-2 approximation. The sum rules predict these form factors 
at large recoil and  small mass of the dipion state.
Second, based on this calculation, we investigate the role of higher partial waves 
in the $B\to \pi\pi$ form factors and assess  the impact of the 
contributions  beyond the $\rho$-meson in the lowest $P$-wave.  
In what follows, the derivation of LCSRs for $B\to \pi \pi$ 
form factors is presented in Sect.~2. In Sect.~3 we compare our
predictions with the $B\to \rho$ form factors. In Sect.~4 using the available 
information on the chiral-odd dipion DA, we calculate the form factors numerically. 
Our conclusions  are presented in Sect.~5. The Appendices contain some details (A)
on the decay kinematics and (B) on the dipion DAs.

\section{Light-cone sum rules with dipion distribution amplitudes}

The LCSR   derivation starts from 
defining an appropriate correlation function. We consider the 
$T$-product of the $b\to u$ 
weak current $j^{V-A}_\mu(x)=\bar{u}(x)\gamma_\mu(1-\gamma_5)b(x)$  
with the $B$-meson interpolating current 
$j_5^{(B)}(0)=im_b\bar{b}(0)\gamma_5d(0)$. Since we are interested in the  
final state with two pions, this $T$-product is then  
sandwiched between the vacuum and the on-shell dipion state:
\begin{equation}
\begin{aligned}
\Pi_\mu(q,k_1,k_2)= i\int\,d^4x e^{iqx}
\langle \pi^+(k_1)\pi^0(k_2)|
T\{j^{V-A}_\mu(x),j_5(0)\} |0\rangle\,.
\end{aligned}
\label{eq:corr}
\end{equation} 
The above correlation function has a more complicated kinematics 
than in the case of the one-pion final state and depends on three independent 
4-momenta $q, k_1,k_2$. 
We denote by $k=k_1+k_2$ the total dipion four-momentum  and by $p=q+k$ 
the external four-momentum of the $B$-meson interpolating current. 
At fixed $k_{1,2}^2=m_\pi^2$ these momenta form  four independent  
invariant variables, as such we 
choose $p^2=(q+k)^2$, $q^2$, $k^2$ and $q\cdot \overline{k}$, where $\overline{k}=k_1-k_2$.
Further details on the kinematics are given in the Appendix A.

The correlation function (\ref{eq:corr}) is decomposed in  four independent 
Lorentz-vectors \footnote{Here we use the convention $\epsilon^{0123}=-1$.}:
\begin{eqnarray}
\Pi_\mu(q,k_1,k_2)=
i\epsilon_{\mu\alpha\beta\rho}q^\alpha k_1^\beta k_2^\rho\, \Pi^{(V)}+
q_\mu\Pi^{(A,q)}
+k_\mu\Pi^{(A,k)}+\overline{k}_\mu\Pi^{(A,\overline{k})}\,,
\label{eq:corr2}
\end{eqnarray} 
where the first term (the rest) corresponds to the contribution of the vector
(axial) part of the  $b\to u$ weak current and  
the invariant amplitudes $\Pi^{(V),(A,q),...}$
depend on the four  invariant variables: $p^2, q^2,k^2,q\cdot\overline{k}$.

To guarantee the validity of the operator-product expansion (OPE) for the 
correlation function (\ref{eq:corr}) near the 
light-cone ($x^2\sim 0$), we consider the region $p^2\ll m_b^2$ and $q^2\ll m_b^2$,
so that the $b$-quark mass provides the large scale. 
In this respect, the conditions for the light-cone dominance  
are practically the same as in the case of the vacuum-to-pion
correlation functions used to obtain the LCSRs for $B\to \pi$ form factors
(for a detailed  derivation of the latter sum rules see, e.g., Ref.~\cite{Duplancic}).
An additional constraint concerns the invariant mass of dipion
which is also kept small, $k^2\lesssim 1 \GeV^2\ll m_b^2$. In this region 
the two-pion system with isospin one is dominated  by the $\rho(770)$ resonance,
accompanied by a nonresonant background. In this paper, we only 
consider the charged dipion state, so that only 
odd angular momenta contribute in the isospin  symmetry  limit. 
This limitation simplifies our analysis, whereas 
the case of neutral dipion state where also the scalar/isoscalar  
$f^0$ resonances contribute, will be considered elsewhere.     

\begin{figure}
\begin{center}
\includegraphics[scale=0.4]{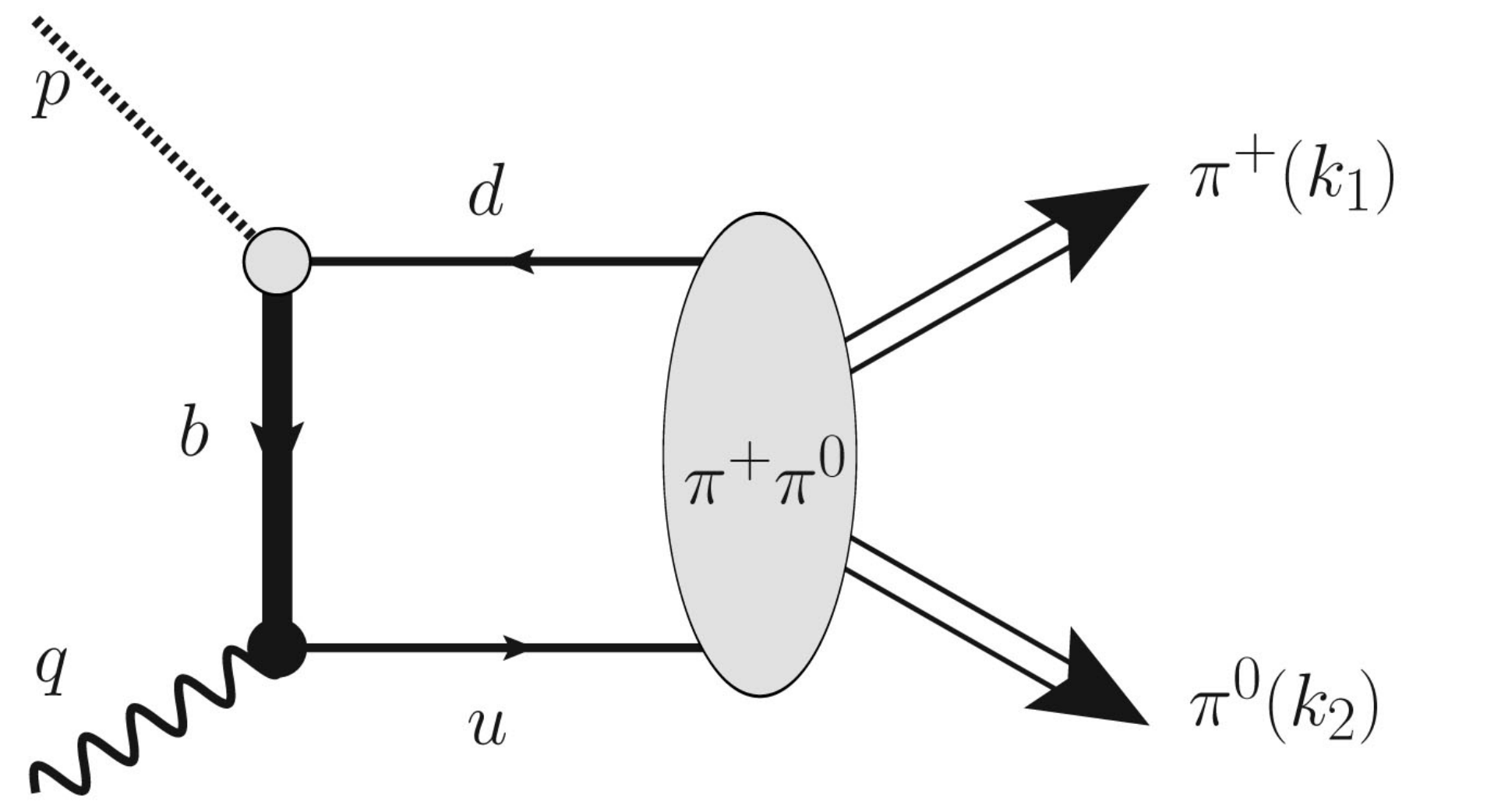}
\end{center}
\caption{ \it Diagram representing the correlation function (\ref{eq:corr}) in leading order;
the wavy (dotted) line represents the weak ($B$-meson interpolating) quark current.}
\label{fig1}
\end{figure}

Turning to the calculation of the correlation function (\ref{eq:corr}), 
in the leading-order (LO) approximation ($\alpha_s=0$), after inserting 
the free $b$-quark propagator, we obtain:
\begin{eqnarray}
\Pi_\mu(q,k_1,k_2)=i\!\!\int \!\!d^4x\!\! \int\!\! \frac{d^4f}{(2\pi)^4}e^{i(q-f)x}\frac{m_b}{m_b^2-f^2}
\langle \pi^+(k_1)\pi^0(k_2)|\bar{u}(x)\gamma_\mu(1-\gamma_5)(\slashed{f}+m_b)\gamma_5d(0)|0\rangle. 
\label{eq:corr3}
\end{eqnarray}
This expression
consists of the  hard-scattering amplitude - the virtual $b$-quark propagator - 
convoluted with  the vacuum $\to$ dipion 
matrix elements of bilocal quark-antiquark  operators. 
These matrix elements absorb long-distance dynamics 
and are expressed via universal dipion DAs, defined following Ref.~\cite{PolyakovNP}. 
The LO diagram of OPE for the correlation function 
(\ref{eq:corr}) is shown in Fig.~\ref{fig1}.

In this paper we will confine ourselves to the leading, twist-2 approximation
for the nonlocal hadronic matrix elements.
We use the following definitions of the twist-2 DAs \cite{PolyakovNP}:
\be
\langle \pi^+(k_1)\pi^0(k_2)|\bar{u}(x)\gamma_\mu[x,0] d(0) |0\rangle=
-\sqrt{2}k_\mu\int\limits_0^1du e^{iu(k\cdot x)}\Phi^{I=1}_{\parallel}(u,\zeta,k^2)\,,
\label{eq:phipar}
\ee
\be
\langle \pi^+(k_1)\pi^0(k_2)|\bar{u}(x)\sigma_{\mu\nu}[x,0] d(0) |0\rangle=
2\sqrt{2}i\frac{k_{1\mu}k_{2\nu}-k_{2\mu}k_{1\nu}}{2\zeta -1}\int\limits_0^1du e^{iu(k\cdot x)}\Phi^{I=1}_{\perp}(u,\zeta,k^2)\,,
\label{eq:phisig}
\ee
where Eq.~(\ref{eq:phipar})  and  Eq.~(\ref{eq:phisig}) represent, respectively, the chiral-even 
 and chiral-odd terms in the light-cone expansion and $[x,0]$ is the gauge factor. 
The DAs depend on the dipion mass squared $k^2$, on the 
fraction $u$ of the two-pion longitudinal momentum carried by the $u$-quark 
(so that $1-u\equiv \bar{u}$ is carried by the $\bar{d}$ quark) and on the 
parameter $\zeta$ related to $q\cdot \overline{k}$ (see Appendix A).
The normalization conditions are \cite{PolyakovNP}:
\be
\int\limits_0^1 du\, \Phi^{I=1}_{\parallel}(u,\zeta,k^2)=(2\zeta-1)F_\pi^{em}(k^2)\,, ~~~
\int\limits_0^1 du\, \Phi^{I=1}_{\perp}(u,\zeta,k^2)=(2\zeta-1)F_\pi^{t}(k^2)\,,
\label{eq:norm}
\ee
where $F_\pi^{em}(k^2)$ is the standard electromagnetic form factor of the pion
in the timelike region (so that $F_\pi^{em}(0)=1$) and $F_\pi^{t}(k^2)$
is the ``tensor'' form factor of the pion normalized to the dimensionful parameter introduced 
in Ref.~\cite{PolyakovNP}:
\be
F_\pi^{t}(0)=1/f_{2\pi}^\perp\,.
\label{eq:norm1}
\ee
The definition (\ref{eq:phipar}) coincides with 
the one introduced in Ref.~\cite{PolyakovNP},
whereas the DA defined in Eq.~(\ref{eq:phisig}) 
differs by the above factor. 
We also use the isospin conventions as
defined in Ref.~\cite{DGP} to relate the dipions  with definite 
isospin projections of pions to the $\langle \pi^+\pi^0|$  state.
Hereafter we omit the isospin index at DAs, since in this paper we 
only consider  the $I=1$ dipion state.
Note that $k^2$ has to be sufficiently small to avoid
large generic  $O(k^2x^2)$ terms in the light-cone expansion. 

In addition to the matrix elements  (\ref{eq:phipar}) and (\ref{eq:phisig}),
one recovers in  Eq.~(\ref{eq:corr3}) also the ones with the 
Dirac matrices $1,\gamma_\mu\gamma_5$; they correspond to the 
higher twists and are neglected here, whereas the nonlocal matrix element  
with $\gamma_5$  vanishes due to $P$-parity conservation.
Sorting out the Dirac structures in  Eq.~(\ref{eq:corr3}) and applying the definitions 
of DAs we obtain, at twist-2 accuracy: 
\begin{eqnarray}
\Pi_\mu(q,k_1,k_2)= i\sqrt{2} m_b\int\limits_0^1 \frac{du}{(q+uk)^2-m_b^2}
\Big\{ \Big[ (q\cdot\overline{k})k_\mu- \bigg((q\cdot k)+uk^2\bigg)\overline{k}_\mu
\nonumber\\
+i\epsilon_{\mu\alpha\beta\rho}q^\alpha k_1^\beta k_2^\rho
\Big]\frac{\Phi_\perp(u,\zeta,k^2)}{2\zeta-1}
-m_bk_\mu\Phi_\parallel(u,\zeta,k^2)
\Big\}\,.
\label{eq:pimu}
\end{eqnarray}
From the above expression one reads off 
the invariant amplitudes $\Pi^{(r)}$ defined in  Eq.~(\ref{eq:corr2}) with 
$(r)= (V),(A,q),(A,k), (A,\overline{k})$, and represents them with a generic expression:
\begin{equation}
\Pi^{(r)}(p^2,q^2,k^2, \zeta)=\sum_{i=\parallel,\perp}\int\limits_0^1 du \frac{f^{(r)}_i(p^2,q^2,k^2,\zeta)
\Phi_i(u,\zeta,k^2) }{(q+uk)^2-m_b^2}\,, 
\label{eq:pi}
\end{equation}
where  the coefficient function convoluted with the dipion DA's
consists of  
the $b$-quark propagator multiplied by a certain 
kinematical factor $f^{(r)}_i$ .  
Transforming the integration variable $u$ to
\begin{equation}
s(u)=\frac{m_b^2-q^2\bar{u}+k^2u\bar{u}}{u}\,,
\label{eq:uvss}
\end{equation}
we bring the integral in   Eq.~(\ref{eq:pi}) to 
a  dispersion form in the variable $p^2$:
\be
\Pi^{(r)}(p^2,q^2,k^2, \zeta)=\sum_{i=\parallel,\perp} f_i^{(r)}(p^2,q^2,k^2,\xi)\int\limits_{m_b^2}^\infty
\frac{ds}{s-p^2} \left (\frac{du}{ds}\right) \Phi_i(u(s),\zeta,k^2)\,.
\label{eq:sint}
\ee
The coefficient functions in the above, after transforming the variable:  
$f_i^{(r)}(p^2,q^2,k^2,\xi)= f_i^{(r)}(p^2-s+s,q^2,k^2,\xi)$, can be
expanded in the powers of $(p^2-s)$, which will vanish 
after the Borel transformation of  Eq.~(\ref{eq:sint}) in $p^2$  
used below. Hence, we can simply replace 
$p^2\to s$ in  Eq.~(\ref{eq:sint}) 
and put the functions $f_i^{(r)}(s,q^2,k^2,\xi)$  under the integral,
as a part of the spectral density.
However, due to a more complicated  kinematics of the correlation 
function, this replacement is not legitimate 
in one particular invariant amplitude multiplying $k_\mu$.  
In this case the function $f_i^{(A,k)}(p^2,q^2,k^2,\xi)$ contains 
the factor ${q\cdot\overline{k}}=1/2(2\xi-1)\lambda^{1/2}(p^2,q^2,k^2)$
(see Appendix A for details). This factor, after analytical continuation 
in $p^2$, generates  a cut at the real axis, more specifically at   
$(\sqrt{q^2}-\sqrt{k^2})^2< p^2<(\sqrt{q^2}+\sqrt{k^2})^2$, which 
does not correspond to any physical intermediate state and represents
a typical kinematic singularity.  Moreover, after Borel transformation, 
the contribution of this cut to the dispersion integral 
is enhanced with respect to the  $b$-quark spectral density. 
Hence, within the framework
of the standard sum rule procedure, we are only in a position to 
derive the LCSRs for the invariant amplitudes $\Pi^{(V)}$ and $\Pi^{(A,\overline{k})}$.

The derivation of these LCSRs continues along the same lines as in the 
well-known case of $B\to \pi$ form factor (see e.g., Ref.~\cite{Duplancic}).  
Applying the quark-hadron duality approximation,
one introduces the effective threshold  $s^B_0$ in the $B$-meson channel,
so that the part of the integral in  Eq.~(\ref{eq:sint}) from $s^B_0$ to $\infty$ 
is approximated by its duality-counterpart in the hadronic 
dispersion relation and subtracted. 
After that, the Borel transformation with respect to the variable 
$p^2\to M^2$ is applied. The result in generic form is:
\be
\Pi^{(r)}(M^2,s_0^B,q^2,k^2, \zeta)= 
\sum_{i=\parallel,\perp} \int\limits_{m_b^2}^{s^B_0} ds ~e^{-s/M^2}
f_i^{(r)}(s,q^2,k^2,\xi) \frac{du}{ds} \Phi_i(u(s),\zeta,k^2)\,,
\label{eq:sint1}
\ee
At this stage it is convenient  to return to the original integration
variable $u$, using the inverse transformation of  Eq.~(\ref{eq:uvss}):
\be  
u(s)=\frac{k^2+q^2-s+\sqrt{4k^2(m_b^2-q^2)+(s-k^2-q^2)^2}}{2k^2}\,.
\label{svsu}
\ee
The following expressions  for the 
Borel-transformed and subtracted invariant amplitudes are obtained:
\begin{eqnarray}
\Pi^{(V)}(M^2,s_0^B,q^2,k^2, \zeta)=
-\frac{2 \sqrt{2} i m_b}{2\zeta-1}\int\limits_{u_0}^1\frac{du}{u}
e^{-\frac{m_b^2-q^2\bar{u}+k^2u\bar{u}}{uM^2}}\Phi_{\perp}(u,\zeta,k^2)\, ,
\label{eq:PiV}\\
\Pi^{(A,\overline{k})}(M^2,s_0^B,q^2,k^2, \zeta)=
\frac{\sqrt{2}im_b}{2(2\zeta-1)}\int\limits_{u_0}^1\frac{du}{u^2}
e^{-\frac{m_b^2-q^2\bar{u}+k^2u\bar{u}}{uM^2}}\Big(m_b^2-q^2+k^2u^2)\Big)
\Phi_{\perp}(u,\zeta,k^2)\,,
\label{eq:invampl}
\end{eqnarray}
where $u_0=u(s_0)$.
In addition, the condition: 
\be
\Pi^{(A,q)}(M^2,s_0^B,q^2,k^2, \zeta)=\,0\,,
\label{eq:Piq}
\ee
is valid at the twist-2 order.

To proceed,
we use  the hadronic dispersion relation for the correlation function 
in the variable
$p^2$ where we only retain the ground $B$-meson state contribution:
\be
\Pi_\mu(q,k_1,k_2)=\frac{\langle \pi^+(k_1)\pi^0(k_2)|\bar{u}\gamma_\mu(1-\gamma_5)b|
\bar{B}^0(p)\rangle f_B m_B^2}{m_B^2-p^2}+.... \,,
\label{eq:hadr_disp}
\ee
with the decay constant of $B$-meson defined via
$\langle \bar{B}^0(p)| \bar{b}\,im_b\gamma_5 d|0\rangle=f_Bm_B^2$.
In  Eq.~(\ref{eq:hadr_disp}) the ellipses denote the contributions 
of radially excited and continuum  states with $B$-meson quantum numbers,
approximated employing the quark-hadron duality approximation.

We then decompose the $B\to \pi\pi$ transition matrix element in the 
form factors we are interested in. We use the  
definition similar to the one in Ref.~\cite{FFKMvD}
\footnote{ 
Our definition of the form factors differs from the one in Ref.~\cite{FFKMvD}
only by some phase factors, caused by a difference in 
the conventions for the $\epsilon$-tensor and for the 
phase of the dipion state.}:
\begin{eqnarray}
 i\langle \pi^+(k_1) \pi^0(k_2)|\bar{u}\gamma^\mu(1-\gamma_5)b|\bar{B}^0(p)\rangle
    = -F_\perp \, \frac{4}{\sqrt{k^2 \lambda_B}} \, 
i\epsilon^{\mu\alpha\beta\gamma} \, q_\alpha \, k_{1\beta} \, k_{2\gamma}
\nonumber \\
    + F_t \, \frac{q^\mu}{\sqrt{q^2}} 
    + F_0 \, \frac{2\sqrt{q^2}}{\sqrt{\lambda_B}} \,
\Big(k^\mu - \frac{k \cdot q}{q^2} q^\mu\Big)
\nonumber\\
    + F_\parallel \, \frac{1}{\sqrt{k^2}} \, 
 \Big(\overline{k}^\mu - \frac{4 (q\cdot k) (q \cdot \overline{k})}{\lambda_B} \, k^\mu
+ \frac{4 k^2 (q\cdot \overline{k})}{\lambda_B} \, q^\mu \Big)\,,
\label{eq:formf}
\end{eqnarray}
where 
\be
\lambda_B\equiv\lambda(m_B^2,q^2,k^2),~~~
q\cdot k=\frac12(m_B^2-q^2-k^2),~~~ q\cdot \overline k =\frac12(2\zeta-1)\lambda_B\,.
\label{eq:rel-2}
\ee
The $B \to 2\pi$ form factors $F_\perp$  and $F_{t,0,\parallel}$ 
depend on the variables $q^2,k^2$ and $q\cdot \overline{k}$, 
and parametrize the transition  matrix element of the vector and axial 
weak $b\to u$ currents, respectively. 
Hereafter, we replace in the form factors  
the variable $q \cdot \overline k$  by  $\zeta$, 
using the relation (\ref{eq:rel-2}).
The form factors defined in  Eq.~(\ref{eq:formf}) can be expanded in partial waves:
\begin{align}\label{eq:ffexpansionhel1}
F_{0,t}(q^2,k^2,\zeta)
=&\sum\limits_{\ell=0}^{\infty}\sqrt{2 \ell +1} F_{0,t}^{(\ell)}(q^2,k^2) P_\ell^{(0)}(\cos\theta_\pi),
\\
\label{eq:ffexpansionhel2}
F_{\perp,\parallel}(q^2,k^2,\zeta)
=&
\sum\limits_{\ell=1}^{\infty}\sqrt{2 \ell +1} F_{\perp,\parallel}^{(\ell)}(q^2,k^2) \frac{P_\ell^{(1)}(\cos\theta_\pi)}{\sin \theta_\pi}\,,
\end{align}
where $P_l^{(m)}$ are the (associated) Legendre polynomials, and $\theta_\pi$ is the 
angle between the pions  in their c.m. frame, related to the parameter $\zeta$ via:     
\be
(2\zeta -1)= \beta_\pi cos \theta_\pi\,,~~\beta_\pi\equiv \sqrt{1-4m_\pi^2/k^2}\,.
\label{eq:zetaTheta}
\ee
Substituting the decomposition (\ref{eq:formf}) in   Eq.~(\ref{eq:hadr_disp}), we 
match the hadronic dispersion relation to the OPE result for 
the correlation function $\Pi_\mu$.
For each invariant amplitude in the decomposition (\ref{eq:corr2}) 
a separate equation  is obtained relating it to one of the form factors 
or to their linear combination.  For the OPE result after
the subtraction of higher than $B$-meson states 
and Borel transformation,
we can directly use the expressions given in  Eqs.~(\ref{eq:PiV}) and (\ref{eq:invampl}).
For the vector-current form factor we obtain the following LCSR
in the adopted LO and twist-2 approximation: 
\be
 \frac{F_\perp(q^2,k^2,\zeta)}{\sqrt{k^2}\sqrt{\lambda_B}}= -\frac{m_b}{\sqrt{2}f_B m_B^2(2\zeta-1)}
\int\limits_{u_0}^1\frac{du}{u}
\Phi_{\perp}(u,\zeta,k^2)\, e^{\frac{m_B^2}{M^2}-\frac{m_b^2-q^2\bar{u}+k^2u\bar{u}}{uM^2}}
\,.
\label{eq:FperpSR}
\ee
Furthermore, 
equating the coefficients at $\overline{k}_\mu$ in the 
OPE and hadronic representations of the correlation function,
yields the LCSR for the one of the axial-current form factors: 
\begin{eqnarray}
\frac{F_{\parallel}(q^2,k^2,\zeta)}{\sqrt{k^2}}= 
-\frac{m_b}{\sqrt{2}f_Bm_B^2(2\zeta-1)}\int\limits_{u_0}^1\frac{du}{u^2}
\Big(m_b^2-q^2+k^2u^2\Big)
\Phi_{\perp}(u,\zeta,k^2)\,
 e^{\frac{m_B^2}{M^2}-\frac{m_b^2-q^2\bar{u}+k^2u\bar{u}}{uM^2}}\,.
\label{eq:FparSR}
\end{eqnarray}
Finally, since the invariant amplitude multiplying 
$q_\mu$ vanishes, 
an additional  relation between the axial-current form factors emerges: 
\be
F_t(q^2,k^2,\zeta)= 
\frac{1}
{\sqrt{\lambda_B}}\Big[(m_B^2-q^2-k^2)F_0(q^2,k^2,\zeta)-
2\sqrt{k^2}\sqrt{q^2}(2\zeta-1)F_{\parallel}(q^2,k^2,\zeta)\Big]\,.
\label{eq:ffrel}
\ee
Note that the remaining invariant amplitude multiplying $k^\mu$ 
contains irreducible
kinematical singularities mentioned above, hence, the additional sum rule 
which could yield the form factor $F_0$ cannot 
be derived with the same method. Hence, in the following we confine ourselves
by analyzing in detail the LCSRs for the form factors $F_\perp$ and $F_{\parallel}$\,. Interestingly, both sum rules depend on the single, chiral-odd 
dipion DA defined in  Eq.~(\ref{eq:phisig}). 

Following Ref.~\cite{PolyakovNP}, 
we represent this DA  in a form of the   
double expansion  in Legendre and Gegenbauer polynomials:
\begin{align}\label{eq:daexpansion}
\Phi_{\perp}(u,\zeta,k^2)
=&
\frac{6 u(1-u)}{f_{2\pi}^\perp}  \sum\limits_{n=0,2,..}^{\infty}\,\,\sum\limits_{\ell=1,3,..}^{n+1}
B_{n\ell}^{\perp}(k^2) C^{3/2}_n (2u-1) \beta_\pi P_\ell^{(0)}\Big(\frac{2\zeta -1}{\beta_\pi}\Big),
\end{align} 
with multiplicatively renormalizable coefficients
$B_{n\ell}^{\perp}(k^2)$ (see Appendix B for more details).
Note that the index $n$ ($l$) goes
over even (odd) numbers and  the normalization conditions 
(\ref{eq:norm}), (\ref{eq:norm1}) yield for the lowest coefficient 
$B_{01}^{\perp}(0)=1 $.
The coefficients with $n\geq 2$ play the same role as the Gegenbauer
moments of the twist-2 pion DA. The values of $B_{(n\geq 2)\ell}^{\perp}(k^2)$ at a low scale determine the nonasymptotic
part of the DA, logarithmically decreasing at large scales.  
Importantly, if one adopts a certain approximation for the nonasymptotic 
part of DA, that is, truncates the expansion (\ref{eq:daexpansion}) 
at a given $n_{max}$, the values of $\ell$ are restricted to $n_{max}+1$.
The coefficients $B_{nl}^{\perp}(k^2)$ are complex functions of 
the dipion invariant mass,
with the imaginary part  at $k^2>4m_\pi^2$, due to 
the unitarity  relation. 
Note that the function  $B_{01}^{\perp}(k^2)$ 
is reduced to the timelike ``tensor'' form factor of the 
pion, which cannot be simply extracted from experiment.

Furthermore, we substitute the partial wave expansion (\ref{eq:ffexpansionhel2}) 
in l.h.s. and the double expansion (\ref{eq:daexpansion}) in r.h.s.
of the LCSRs (\ref{eq:FperpSR}) and (\ref{eq:FparSR}), replacing in the r.h.s. the argument of the Legendre polynomial by $\cos\theta_\pi$,
according to  Eq.~(\ref{eq:zetaTheta}). Multiplying both parts of the resulting relation 
by $\sin\theta_\pi P_{\ell'}^{(1)}(\cos\theta_\pi)$ and integrating over 
$\cos\theta_\pi$ we use the orthogonality relation:
\be
\int\limits_{-1}^{+1}dz P_{\ell}^{(1)}(z) P_{\ell'}^{(1)}(z)=
\frac{2(\ell+1)!}{(2\ell+1)(\ell-1)!}\delta_{\ell \ell'}
\ee
and obtain the sum rules for the $\ell$-th  partial wave contribution  to the 
$B\to 2\pi$ form factors $(\ell=1,3,...)$:
\ba  
F_{\perp}^{(\ell)}(q^2,k^2)=
\frac{\sqrt{k^2}}{\sqrt{2}f_{2\pi}^\perp}\frac{\sqrt{\lambda_B} m_b}{m_B^2 f_B}
e^{m_B^2/M^2}\sum\limits_{n=0,2,..}\,\,\sum_{\ell'=1,3,..}^{n+1}I_{\ell \ell'}\,B_{n\ell'}^\perp(k^2)
J_n^{\perp}(q^2,k^2,M^2,s_0^B)\,,
\label{eq:lcsr_part_perp}
\ea
\ba  
F_{\parallel}^{(\ell)}(q^2,k^2)=
\frac{\sqrt{k^2}}{\sqrt{2}f_{2\pi}^\perp}\frac{m_b^3}{m_B^2 f_B}
e^{m_B^2/M^2}\sum\limits_{n=0,2,4,..}\,\sum_{\ell'=1,3,..}^{n+1}I_{\ell \ell'}\,B_{n\ell'}^\perp(k^2)
J_n^{\parallel}(q^2,k^2,M^2,s_0^B)\,,
\label{eq:lcsr_part_par}
\ea
where the short-hand notation is introduced for 
the angular integral:
\be
I_{\ell \ell'}\equiv -\frac{\sqrt{2\ell+1}(\ell-1)!}{2 (\ell+1)! }
\int\limits_{-1}^{+1}\frac{dz}{z}\sqrt{1-z^2} P_{\ell}^{(1)}(z)P_{\ell'}^{(0)}(z)\,,
\ee
so that, e.g., $I_{1,1}=1/\sqrt{3}$, $I_{1,3}=-1/\sqrt{3}$, $I_{1,5}=4/(5\sqrt{3})$, 
and the integrals over the quark-momentum fraction are defined as
\be
J_n^{\perp}(q^2,k^2,M^2,s_0^B)= 6\!\int\limits_{u_0}^1\! du(1-u)C_n^{3/2}(2u-1)
e^{-\frac{m_b^2-q^2\bar{u}+k^2u\bar{u}}{uM^2}}\,,
\ee
and
\be
J_n^{\parallel}(q^2,k^2,M^2,s_0^B)= 6\!\int\limits_{u_0}^1\! \frac{du}{u}
(1-u)C_n^{3/2}(2u-1)
\Big(1-\frac{q^2-k^2u^2}{m_b^2}\Big)e^{-\frac{m_b^2-q^2\bar{u}+k^2u\bar{u}}{uM^2}}\,.
\ee
Note that $I_{ll'}=0$ at $\ell>\ell'$, hence, in the limit of the asymptotic DA, 
that is, when all coefficients $B_{n\ell}$, 
except $B_{01}$, vanish,
only the $\ell=1$ (partial $P$-wave) term remains in the form factors.
Altogether, the LCSRs (\ref{eq:lcsr_part_perp}) and (\ref{eq:lcsr_part_par})
allow us to assess the relative importance of the higher partial waves 
with $\ell=3,5,..$ in the $B\to \pi\pi$  form factors. One simply 
has to calculate the ratio:  
\be
R^{(\ell)}_{\perp,\parallel}(q^2,k^2)= 
\frac{F_{\perp,\parallel}^{(\ell)}(q^2,k^2)}{F_{\perp,\parallel}^{(1)}(q^2,k^2)}\,.  
\label{eq:ratiosl}
\ee

\section{ How much $\rho$ the $B\to 2\pi$ form 
factors contain?}
Having at our disposal the LCSR 
calculation of the  
$\bar{B}^0\to \pi^+\pi^0$ form factors, we now address another important question: 
the dominance of the $\rho$-meson contribution to these form factors.
This knowledge is indispensable for an accurate interpretation 
of the $B\to \pi\pi \ell \nu_\ell$ measurements. With more data 
on this decay available in future, the angular analysis can in principle 
isolate the  final-state dipion in the $P$-wave from other partial waves. 
It is then important to clarify if  the events in the interval 
of dipion invariant  mass around the $\rho$-meson mass, 
at $\sqrt{k^2} \sim m_\rho\pm \Gamma^{tot}_\rho/2$, 
originate predominantly from the $B\to \rho$ transition, or there is a
noticeable interference with excited $\rho$ resonances and/or $2\pi$ 
($P$-wave) continuum background.
Strictly speaking, the answer to this question 
relies on a (model-dependent) parametrization of the $\rho$ resonance 
and nonresonant background. An approach to the $B\to \pi\pi$ form factors 
at low dipion masses  that is independent of the resonance model and employs 
the hadronic dispersion relation in the variable $k^2$  was suggested in Ref.~\cite{Kubisetal} where the $\pi\pi$ 
rescattering effects, as well as the effect of the $\rho$ meson,
were taken into account employing the Omn\'es representation and 
the data on the pion scattering phases. 

Within the LCSR framework, a similar approach would correspond 
to using a hadronic dispersion relation for the coefficients 
$B_{nl}^\perp(k^2)$ treated as analytical functions of $k^2$. 
An attempt in this direction was already made in Ref.~\cite{PolyakovNP} where 
these coefficients at  low mass $(k^2>~4m_\pi^2)$ were calculated 
in the instanton
model of QCD vacuum and the Omn\'es representation including 
the $\rho$-resonance effect 
was used to extrapolate them towards $k^2\sim 1 $ GeV$^2$. 
We postpone a more detailed study along these lines 
to a future work.

Here we address a different aspect that has an immediate importance 
for the LCSR approach:
are the $B\to \pi\pi$ form factors predicted from LCSRs at low 
dipion masses $k^2\sim 4m_\pi^2$ 
conform  and/or consistent with the $B\to \rho$ form factors calculated 
from the LCSRs with the $\rho$-meson DAs defined in the zero-width approximation. 
To this end,  we employ the hadronic dispersion relation 
for the $P$-wave ($l=1$) part of the $B\to \pi\pi$ form factors in $k^2$ and 
retain only  the intermediate $\rho$-resonance contribution. A more detailed derivation of these 
relations can be found in Ref.~\cite{FFKMvD}. For the two form factors considered 
above we obtain: 
\be
\frac{\sqrt{3}F_{\perp}^{(\ell=1)}(q^2,k^2)}{\sqrt{k^2}\sqrt{\lambda_B}}=
\frac{g_{\rho\pi\pi}}{m_\rho^2-k^2-im_\rho\Gamma_\rho(k^2)}\frac{V^{B\to \rho}(q^2)}{m_B+m_\rho}
+ ...
\label{eq:VrelFperp}
\ee
and 
\be
\frac{\sqrt{3}F_{\parallel}^{(\ell=1)}(q^2,k^2)}{\sqrt{k^2}}=
\frac{g_{\rho\pi\pi}}{m_\rho^2-k^2-im_\rho\Gamma_\rho(k^2)} (m_B+m_\rho)A_1^{B\to \rho}(q^2)
+ ...
\label{eq:VrelFpar}
\ee
where the ellipses denote the contributions of excited  states such as $\rho(1450)$
as well as the possible subtraction terms.
Note that here we prefer to use dispersion relations for the complete invariant 
amplitudes multiplying  
the four-momenta in the Lorentz-decomposition (\ref{eq:formf})
of the $B\to \pi\pi$ matrix element\footnote{ Our choice is  
similar  to the standard form-factor decomposition for 
$K_{e4}$ decay (see e.g., Ref.~\cite{daphne}).}, treating these amplitudes
as analytical functions of $k^2$ and avoiding unnecessary kinematic
singularities. To make the $\rho$-resonance description complete, in  
Eqs.~(\ref{eq:VrelFperp}),(\ref{eq:VrelFpar})
an energy dependent total width is added,
defined as:
\be
\Gamma_\rho(k^2)= \frac{m_\rho^2}{k^2}
\left(\frac{k^2-4m_\pi^2}{m_\rho^2-4m_\pi^2}\right)^{3/2}\theta(k^2-4m_\pi^2)
\Gamma^{tot}_\rho\,,
\label{eq:gtot}
\ee  
(see e.g., the discussion in Ref.~\cite{Bruch:2004py}), 
however it does not  play a role at $k^2\sim 4m_\pi^2$.  
The residues of the $\rho$-pole in the dispersion relations (\ref{eq:VrelFperp})
and (\ref{eq:VrelFpar}) contain the $\rho\to 2 \pi$ strong coupling
defined as
$\langle \pi^+(k_1)\pi^0(k_2)|\rho^+(k)\rangle= -g_{\rho\pi\pi}\epsilon^{(\rho)}\cdot (k_1-k_2)$,
($ \epsilon^{(\rho)}$ is the polarization vector of $\rho$ meson) 
and the
$B\to \rho$ form factors 
$V^{B\to \rho}(q^2)$ and $A_1^{B\to \rho}(q^2)$. 
For the latter we use the standard definition: 
\be
\langle \rho^+(k) |\bar{u}\gamma_\mu(1-\gamma_5) b |\bar{B}^0(p)\rangle=\epsilon_{\mu\alpha\beta\gamma}
\epsilon^{*(\rho)}_\alpha p^\beta k^\gamma\frac{2V^{B\to \rho}(q^2)}{m_B+m_\rho}
-i\epsilon^{*(\rho)}_\mu (m_B+m_\rho)A_1^{B\to\rho}(q^2)+ ...\,
\label{eq:BrhoFFdef}
\ee
where ellipses denote the remaining form factors  related
to the axial current. The above decomposition is the same   
as e.g., in Ref.~\cite{BBFFBrho}. There one can also find a detailed derivation of LCSRs
for these form factors in terms of the $\rho$-meson DAs in the same, leading  
twist-2 approximation: 
\be
V^{B\to \rho}(q^2)= \frac{(m_B+m_\rho)m_b}{2m_B^2 f_B}
f_\rho^\perp e^{\frac{m_B^2}{M^2}}
\int\limits_{u_0}^1 \frac{du}{u}~\phi_\perp^{(\rho)}(u)\,
e^{-\frac{m_b^2-q^2\bar{u}+m_\rho^2 u\bar{u}}{uM^2}}\,,
\label{eq:srBrhoV}
\ee
\be
A_1^{B\to \rho}(q^2)= \frac{m_b^3}{2(m_B+m_\rho)m_B^2 f_B}
f_\rho^\perp e^{\frac{m_B^2}{M^2}}
\int\limits_{u_0}^1 \frac{du}{u^2}~\phi_\perp^{(\rho)}(u)
\bigg(1-\frac{q^2-m_\rho^2u^2}{m_b^2}\bigg)
e^{-\frac{m_b^2-q^2\bar{u}+m_\rho^2 u\bar{u}}{uM^2}}\,.
\label{eq:srBrhoA1}
\ee
Note that both sum rules are also determined by the 
chiral-odd DA defined via vacuum $\to \rho$  hadronic matrix element:
\be  
\langle \rho^+(k)| \bar{u}(x)\sigma_{\mu\nu}[x,0] d(0)| 0\rangle=
-i f_\rho^\perp \big(\epsilon^{*(\rho)}_\mu k_\nu -k_\mu\epsilon^{*(\rho)}_\nu \big)
\int\limits_0^1 du e^{iu k\cdot x} \phi_\perp^{(\rho)}(u)\,,
\label{eq:phi_rho}
\ee
and having the Gegenbauer polynomial expansion:
\be
\phi_\perp^{(\rho)}(u)=6u(1-u)\left( 1+\sum\limits_{n=2,4,,..} a_n^{(\rho)\perp} C_n^{3/2}(2u-1)\right)\,,
\label{eq:phirho}
\ee
where the coefficients $a_n^{(\rho)\perp}$ have the same scale-dependence
as the ones in the Gegenbauer expansion of the dipion chiral-odd DA.

The simplest and rather straightforward way to assess the dominance of 
the $\rho$-meson contribution to r.h.s.
of the dispersion relations (\ref{eq:VrelFperp})
and (\ref{eq:VrelFpar}) is to compare numerically both parts of these relations
at $k^2\sim 4m_\pi^2$ where we can evaluate the l.h.s. knowing the coefficients
$B_{nl}^\perp(k^2)$ at low dipion masses. A noticeable difference 
between both sides of these relations will clearly indicate the importance of the heavier 
than $\rho$ states and/or continuum nonresonant background.  
The known higher-twist contributions and gluon radiative corrections  
to the sum rules (\ref{eq:srBrhoV})  and (\ref{eq:srBrhoA1})  (see e.g., Ref.~\cite{BZBV}), 
can be added in future if also the corresponding contributions in the LCSRs for 
$B\to 2 \pi$ form factors are worked out.
   
\section{Numerical analysis}

To specify the numerical input for the LCSRs (\ref{eq:lcsr_part_perp})
and (\ref{eq:lcsr_part_par}), first of all 
we have to adopt a quantitative ansatz for the dipion DAs. This task 
is more complicated than for the single-pion  or $\rho$-meson DAs, 
because the coefficients $B_{nl}^\perp(k^2)$ are now complex functions 
of dipion invariant mass.  More is known on the functions 
$B_{nl}^\parallel(k^2)$, for which the lowest (``asymptotic'') 
one is directly related to the well measured 
pion form factor  in the timelike region: 
$B_{01}^{\parallel}(k^2)=F^{em}_\pi(k^2)$. In addition, some relations 
between $B_{nl}^\parallel(k^2)$  and the Gegenbauer moments of the single-pion DAs are available \cite{PolyakovNP} via soft-pion limit at $k^2\to 0$.
The only available information on the coefficients 
$B_{nl}^\perp(k^2)$ 
are the estimates at low $k^2$  based on the instanton model 
of QCD vacuum \cite{PolyakovNP, PolyakovWeiss}, up to $n=4$.  
We list them  in the Appendix B. For the $\rho$-meson DA we use 
the same ansatz  as the one used in
Ref.~\cite{BBFFBrho}: $a_2^\perp=0.2\pm 0.1$, $a_{n>2}=0$ and  
$f_{\rho}^\perp=160\pm 10 ~{\mbox MeV}$.   

The rest of the input parameters entering LCSRs concerns: (a) the short-distance 
part of the correlation function, (b) the $B$-meson decay  constant  
and (c) the quark-hadron duality approximation for the $B$-meson channel.
In the following we comment on these points:
 
(a) Although here the correlation function 
is known only at LO, and the choice of the renormalization scale cannot be optimized 
without gluon  radiative corrections, in anticipation 
of the future NLO improvement, we adopt the same default scale
$\mu=3$ GeV for all scale-dependent parameters 
including the ones in DAs, following 
the analyses of LCSRs for the $B\to \pi$ form factor in Refs.~\cite{Duplancic,KMOW}. 
We also use the $b$-quark mass 
in $\overline{MS}$ scheme  $\bar{m}_b(\bar{m}_b)=4.18\pm 0.03$ GeV \cite{PDG} and 
adopt the central value $m_b= \bar{m}_b(3~ \mbox{GeV})= 4.47$ GeV, neglecting 
a small uncertainty.  

(b) The two-point 
QCD sum rule for $f_B$ at LO is used, 
which is consistent with our approximation for the LCSRs,
schematically:
\be
f_B^2=[f_B^2]_{2ptSR}(m_b, \langle\bar{q}q\rangle,..., \mu ,\bar{M}^2, \bar{s}_0^B)\,,
\label{eq:fBSR}
\ee
where the  ellipses indicate the vacuum condensate densities of higher
dimensions.
The expression for this sum rule is well known, hence,  
for brevity we do not repeat it here; the values 
of vacuum condensate densities and other parameters are taken the same as in 
the recent analysis \cite{Gelh} (see Table I there).
In particular, we use: for the quark condensate density 
$\langle \bar q q \rangle (2 ~\mbox{GeV})= (-277~\mbox{MeV})^3$, for   
the optimal Borel parameter $\overline{M}\,^2=5.5$ 
GeV$^2$ and for the effective
threshold $\bar{s}_0^B= 34.0 $ GeV$^2$, chosen to reproduce the mass 
of $B$-meson from the sum rule.

(c) We anticipate that the typical Borel parameter values for 
a low dipion mass are in the same ballpark as for the LCSRs 
for the $B\to \pi$ or $B\to \rho$ form factors. For definiteness
we take the interval $M^2=16.0\pm 4.0 $ GeV$^2$  and the corresponding 
threshold values $s_0^B=37.5 \pm 2.5$ GeV$^2$ from the analysis in Ref.~\cite{Duplancic,KMOW}. We expect also that 
LCSRs with dipion DAs are valid in the same region as the 
conventional LCSRs with DAs of single hadron, that is 
at $0\leq q^2\lesssim 12$ GeV$^2$.   
\begin{figure}[t]
{\scriptsize {}\hspace{1.3cm}$F_\perp^{(\ell=1)}(q^2,k^2_{min})$}\hspace{5.7cm}{\scriptsize $F_\parallel^{(\ell=1)}(q^2,k^2_{min})$\hspace{5cm}{}}\\
\vspace{-3mm}
\begin{center}
\includegraphics[scale=0.6]{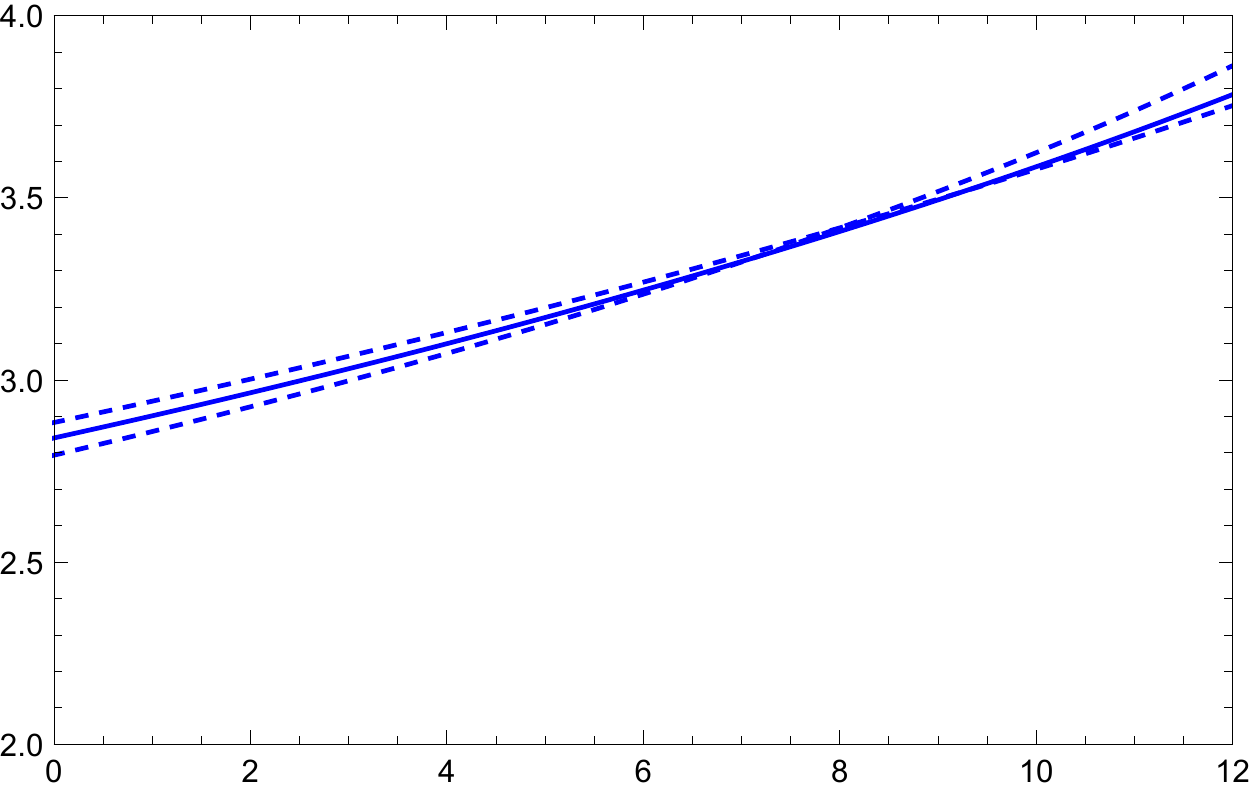}~
\includegraphics[scale=0.6]{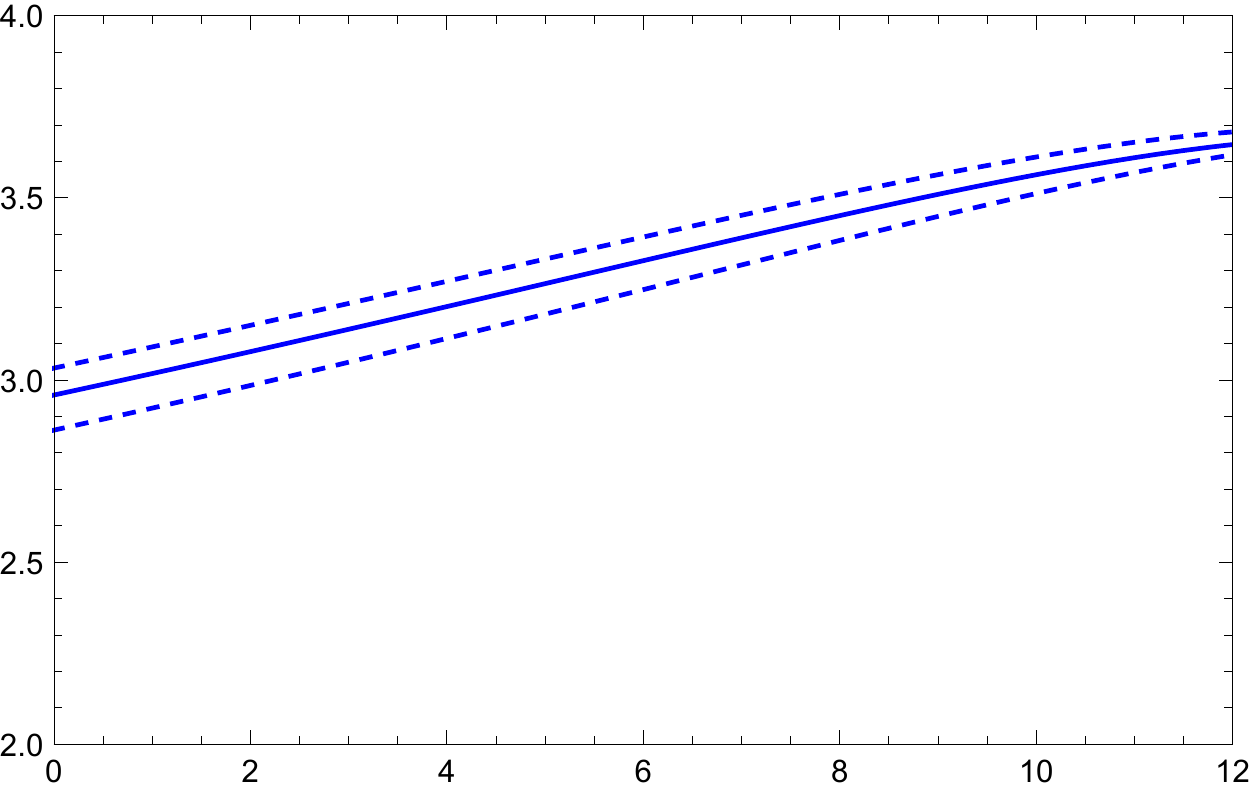}\\
\vspace{-2mm}
{}\hspace{5.9cm}{\scriptsize  $q^2[{\mbox GeV}^2]$}
\hspace{6.6cm}{\scriptsize  $q^2[{\mbox GeV}^2]$}
\end{center}
\caption{\it $P$-wave contributions to $B\to \pi^+ \pi^0$ form factors,
$F_{\perp}^{(\ell=1)}(q^2, k^2_{min})$ (left panel) and 
$F_{\parallel}^{(\ell=1)}(q^2, k^2_{min})$ (right panel), 
calculated from LCSRs
at central values of the input. 
Dashed lines indicate the uncertainty
due to the variation of the Borel parameter.
}
\label{fig2}
\end{figure}

Note that the above input will only serve for numerical illustration
and we postpone the overall analysis of uncertainties, having in mind 
the lack of precision in the new sum rules. Only the Borel-mass 
dependence will be shown for an assessment of the typical sum rule
uncertainties. On the other hand, in 
all ratios of LCSRs used below, the parametrical uncertainties  
are expected to be smaller than in the individual sum rules, 
due to mutual correlations.

\begin{figure}[t]
{\scriptsize 
{}\hspace{1.3cm}
$R_\perp^{(\ell=3)}(q^2,k^2_{min})$}
\hspace{5.4cm}{\scriptsize $R_\parallel^{(\ell=3)}(q^2,k^2_{min})$\hspace{6.3cm}{}}\\
\vspace{-3mm}
\begin{center}
\includegraphics[scale=0.6]{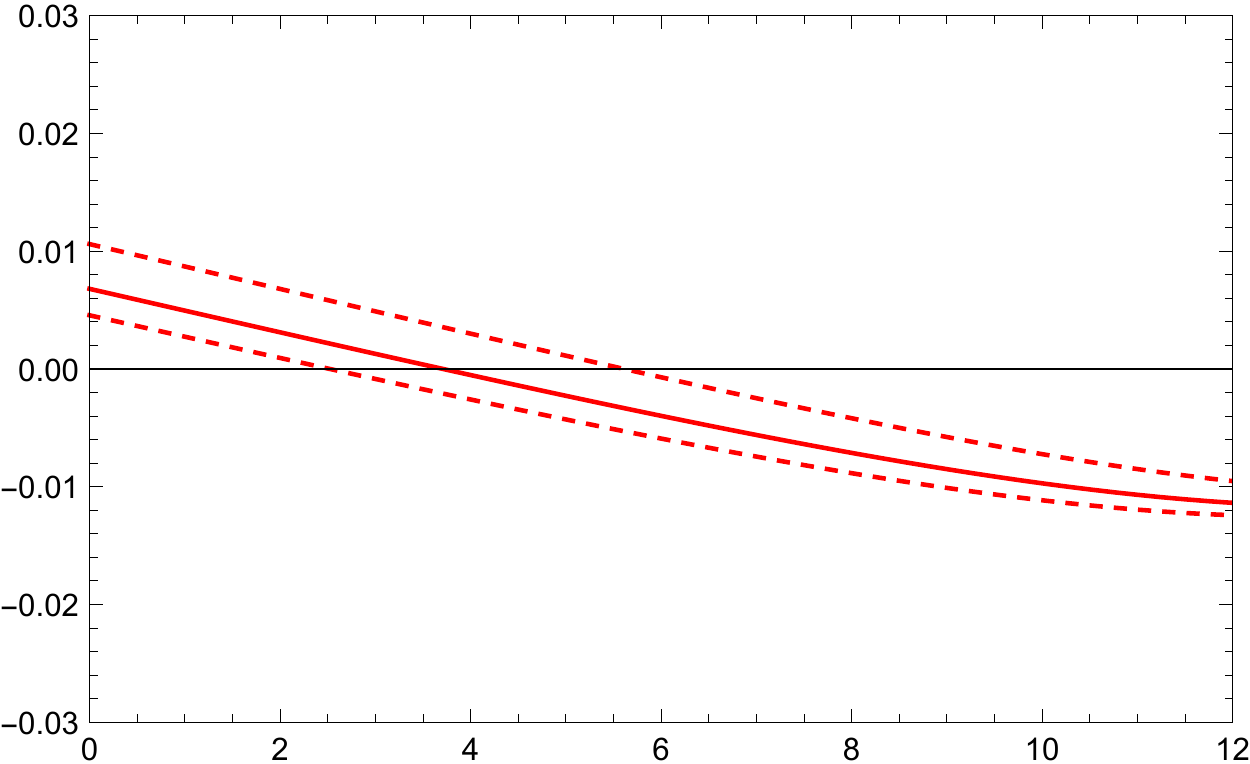}~
\includegraphics[scale=0.6]{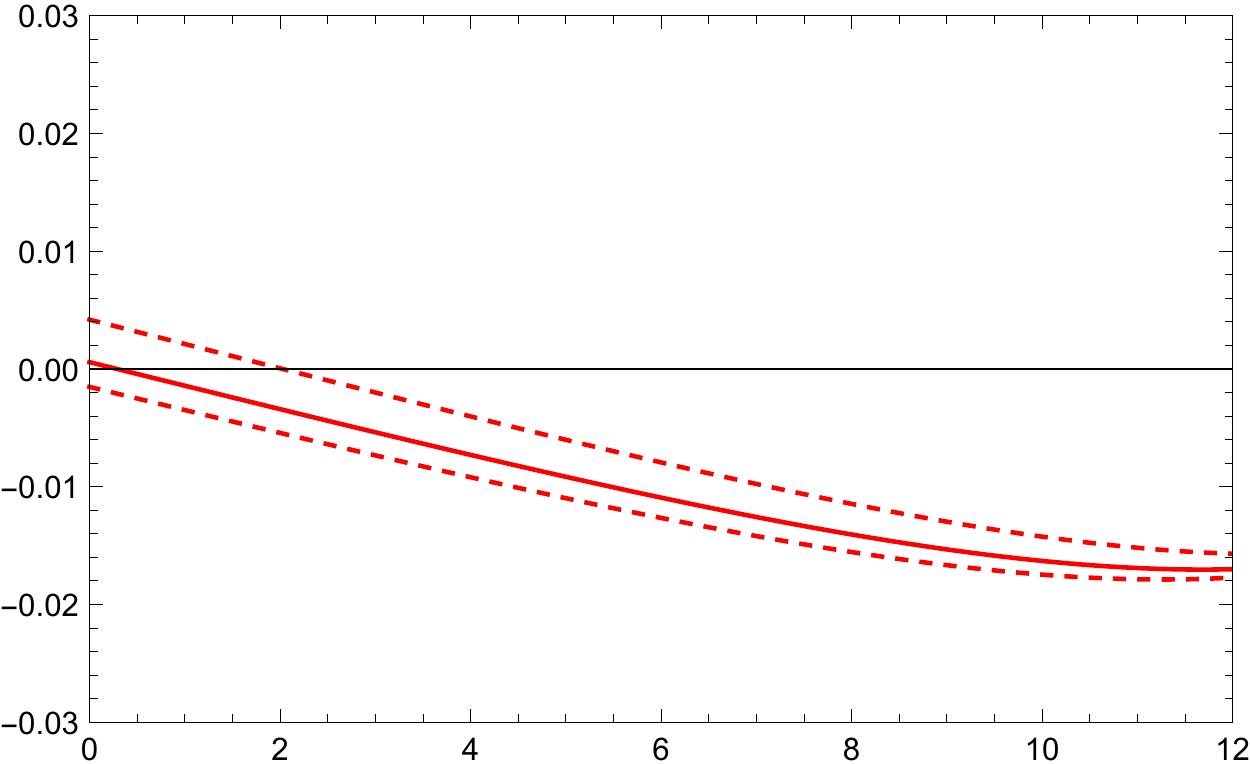}\\
\vspace{-2mm}
{}\hspace{5.9cm}{\scriptsize  $q^2[{\mbox GeV}^2]$}
\hspace{6.6cm}{\scriptsize  $q^2[{\mbox GeV}^2]$}
\end{center}
\caption{\it The ratio of $F$-wave and $P$-wave contributions to 
$B\to \pi^+ \pi^0$ form factors $F_{\perp}$
(left panel) and $F_{\parallel}$ (right panel)
calculated from LCSRs at central values of the input. 
Dashed lines indicate the uncertainty
due to the variation of the Borel parameter.}
\label{fig3}
\end{figure}

Inserting the adopted input in the LCSRs (\ref{eq:lcsr_part_perp})
and (\ref{eq:lcsr_part_par}), we calculate  first 
the numerical results  for the $P$-wave contribution 
$F_{\perp}^{(\ell=1)}(q^2, k^2_{min})$ and $F_{\parallel}^{(\ell)}(q^2, k^2_{min})$
at $k_{min}^2= 4m_\pi^2$ and at  $q^2=0-12.0 $ GeV$^2$.  
They are shown in Fig.\ref{fig2}.
In Fig.~\ref{fig3}   the ratios (\ref{eq:ratiosl}) of $F$-wave ($l=3$) and 
$P$ wave form factors are displayed  as a function of $q^2$.
We realize that in the adopted approximation the LCSRs predicts a very small
contribution of the higher partial waves in both form factors. 
The missing 
higher-twist effects 
\footnote{In fact, one has to mention that the twist 3,4 effects in $B\to \rho$ form factors 
are rather small, at the level of a few percent as, for example, found in 
Ref.~\cite{Ball:1998kk} (see discussion and Fig.5 there in which
the contributions of various twists to the LCSR for $A^{B\to \rho}_1$ form factor  
are plotted). The situation there is markedly different from the LCSRs for $B\to \pi$
form factors where the twist-3 part is strongly enhanced by the normalization 
parameter  $m_\pi^2/(m_u+m_d)$.} 
and NLO corrections as well a more elaborated ansatz for 
the Gegenbauer coefficients $B_{n\ell}^\perp$ can change this ratio, but probably not 
its order of magnitude.

\begin{figure}[t]
{\scriptsize {}\hspace{1.05cm}
$[F_\perp^{(\ell=1)}(q^2,k^2_{min})]^{(\rho)}/
[F_\perp^{(\ell=1)}(q^2,k^2_{min})]^{(LCSR)}$
\hspace{1.6cm} $[F_\parallel^{(\ell=1)}(q^2,k^2_{min})]^{(\rho)}/
[F_\parallel^{(\ell=1)}(q^2,k^2_{min})]^{(LCSR)}$ 
}\\
\vspace{-6mm}
\begin{center}
\includegraphics[scale=0.6]{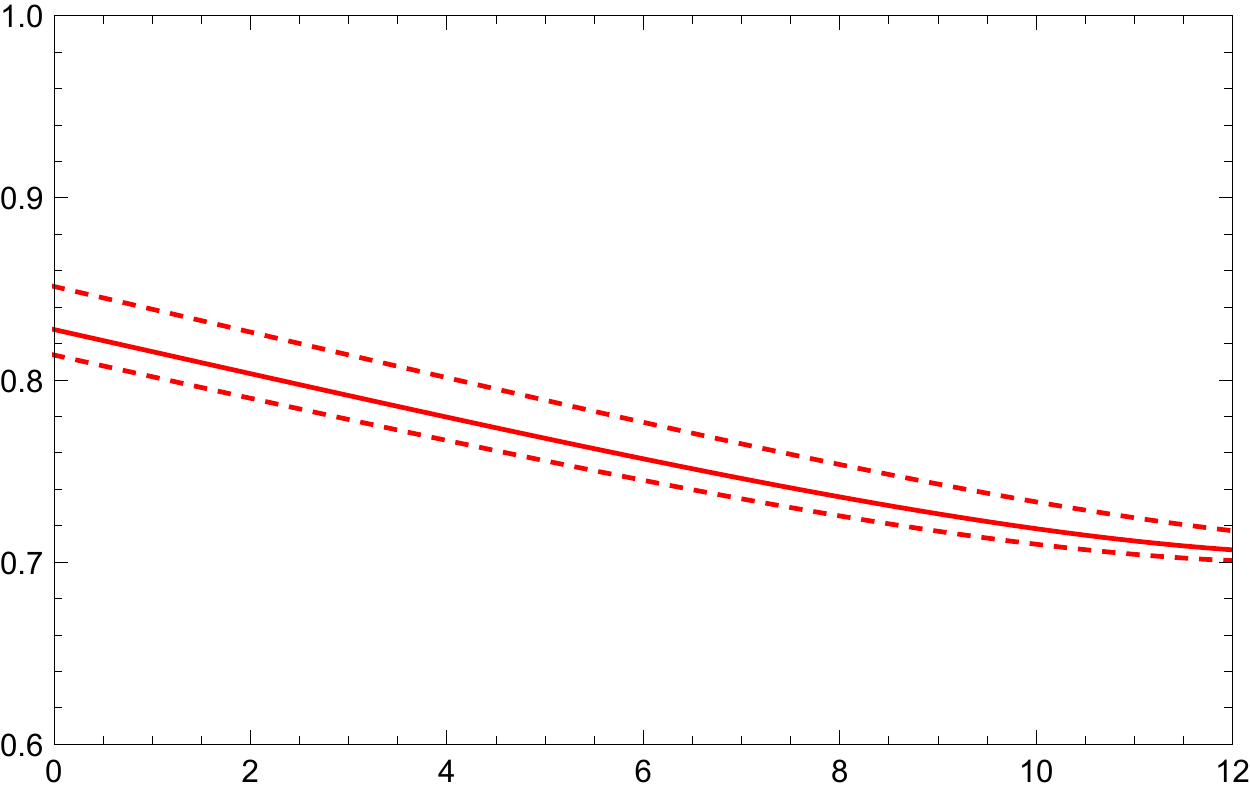}~
\includegraphics[scale=0.6]{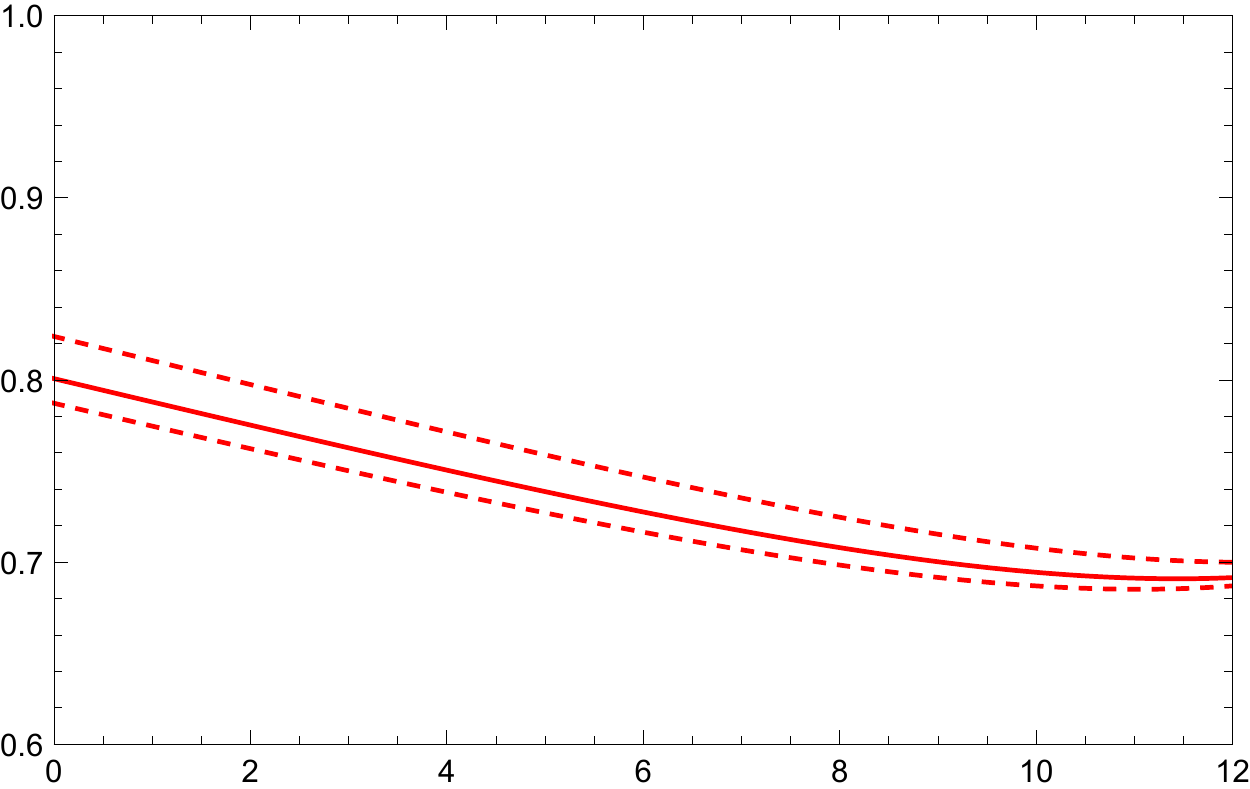}\\
\vspace{-2mm}
{}\hspace{5.9cm}{\scriptsize  $q^2[{\mbox GeV}^2]$}
\hspace{6.6cm}{\scriptsize  $q^2[{\mbox GeV}^2]$}
\end{center}
\caption{\it The relative contribution of $\rho$-meson
to the $P$-wave $B\to \pi^+ \pi^0$  form factors 
$F_\perp^{(\ell=1)}(q^2,k^2_{min})$ (left panel) 
and  $F_\parallel^{(\ell=1)}(q^2,k^2_{min})$ (right panel)  
calculated from LCSRs at central values of the input. 
Dashed lines indicate the uncertainty
due to the variation of the Borel parameter.}
\label{fig4}
\end{figure}

Finally, in Fig.~\ref{fig4} we plot the ratios obtained dividing  
the $\rho$-meson contributions on r.h.s. of  Eqs.~(\ref{eq:VrelFperp}) and (\ref{eq:VrelFpar}),
by the LCSR results for l.h.s of these relations. 
As we see, there is up to  20-30\% ``deficit'' 
which has to be covered by other than $\rho$ contributions to the dispersion relations
for the $B\to \pi\pi$ form factors. A more detailed identification of these 
contributions demands a dispersion relation analysis of DAs in the LCSRs 
as already mentioned above.

\section{Conclusion}
In this paper we presented the first systematic derivation of LCSRs
for the form factors of $B\to \pi\pi$ semileptonic transitions 
in terms of dipion light-cone DAs. We considered the case 
with an odd angular-momentum (isospin one) dipion state, 
so that the dependence on the angle $\theta_\pi$ 
(or equivalently on the invariant variable $q\cdot \bar k$) becomes essential.
As we have shown, the presence of this variable  complicates the 
derivation of sum rules, producing 
in separate cases kinematical singularities in the underlying 
correlation function. 
We concentrated on two particular form factors for which 
the sum rules are free from ambiguities. In the twist-2 approximation,
the resulting LCSRs are determined by a single, chiral-odd dipion DA. 
We obtained numerical predictions at small $k^2$ employing 
the available nonperturbative estimate of the coefficients
in the expansion for this DA. 

Apart from the two sum rules for the $F_\perp$ and $F_{\parallel}$ form factors,  
we also  found a relation between two remaining $\bar{B}^0\to \pi^+\pi^0$ form factors 
$F_{0}$ and $F_{t}$ in twist-2 approximation. The remaining question 
is: how to circumvent the problem of kinematical singularities
and derive an additional LCSR  for one of the latter form factors,
in order to be able to predict their full set. One possibility, 
a subject of a future 
investigation, is to modify the correlation function, e.g., by
employing a different interpolating heavy-light current for $B$ meson, so that 
the form factor we need is contained in a kinematical structure 
free from singularities.       

After partial wave expansion, the new sum rules quantify the contributions  of higher partial waves to the 
$B\to \pi\pi$ form factors. These contributions turn out
to be very small with respect to the lowest $P$-wave 
form factors. Furthermore, in the latter, according to LCSRs, 
the dominance  of the $\rho$-meson terms parametrized 
using the LCSRs for $B\to \rho$ form factors is violated at the 
level of 20-30\%. 

The question of $\rho$-meson dominance in the $B\to \pi\pi$ form factors 
was recently discussed in Ref.~\cite{Straub:2015ica} where the LCSRs
for $B\to \rho,K^*$ form factors were updated. 
There it was argued that 
the $\rho$-state  effectively includes the nonresonant 
background in the $P$-wave dipion state in the experimental as well as the LCSR prediction
for $B\to \rho$. Concerning experimental determination of the 
$\rho$-meson decay constant, this statement does not reflect, 
e.g., the most up-to-date experimental analyses 
of $e^+e^-\to 2\pi$ and  $\tau\to \pi \pi \nu_\tau$   
done by CMD-2 \cite{Akhmetshin:2006bx} and Belle \cite{Fujikawa:2008ma} collaborations, respectively. In both cases
the experimentalists use a model of the timelike pion form factor,
explicitly taking into account the excited states, e.g., adding a separate 
$\rho(1450)$-resonance contribution to  the $\rho$-meson contribution
and then fitting the resonance parameters.
In the similar way, one can 
assess the $\rho$-meson
dominance in $B\to \pi\pi$ form factors at  
a quantitative level, including 
the $B\to \rho(1450)$ transition   
in the dispersion relations (\ref{eq:VrelFperp}) 
and (\ref{eq:VrelFpar}), so that in the $k^2\lesssim m_\rho^2$ region 
this contribution represents a nonresonant $B\to \pi\pi$ background
interfering with the $B\to \rho$ contribution. We emphasize that the dominance
of the $\rho$-meson and the shape of the nonresonant background 
are important issues for the $B\to \pi\pi\ell \nu_\ell$  decays. 
They will be addressed in future using   
available LCSR results for the $B\to \rho$ form factors and 
more accurate LCSR analyses of $B\to \pi\pi $ 
form factors. 

In the literature,  
an earlier attempt to use the dipion DAs in 
the LCSRs for $B\to \pi\pi$ form factors can be found  in Ref.~\cite{Maul}.
However, in that analysis an expansion of the correlation function,
including the factor $\lambda^{1/2}(p^2,q^2,k^2)$, in powers of the dipion mass 
$k^2$ was used. We doubt that in the 
presence of kinematical singularities, discussed above, 
such an expansion is
legitimate, also in  the resulting form factors presented in Ref.~\cite{Maul}, 
the most important contribution of the chiral-odd DA was neglected. 

Recently, the LCSRs for $B\to K\pi$ form factors were obtained in Ref.~\cite{WWang}
employing the DAs  of the $K\pi$ system  in the $S$-wave  state,
in this case the  generalized DAs have the same  form as the DAs for a 
light scalar meson, with no dependence on the variable $\zeta$.
In Ref.~\cite{WWang}, the twist-2 and twist-3 contributions are
taken into account and  their common normalization is related to the 
main  input, the scalar $K\pi$ form factor calculated within the 
chiral perturbation theory framework in \cite{Doring:2013wka}. 
This result provides an estimate for the $S$-wave contribution to 
the form factors of the FCNC $B\to K\pi \ell^+\ell^-$ decays.

The calculation presented in our paper can also be extended 
to the dimeson states with strangeness.
If one removes the $S$-wave constraint on the $K\pi$ state chosen in \cite{WWang}, 
it is possible to access the  $B\to K \pi$ transition form factors with a 
kaon-pion state in the $P$-wave and higher partial waves, 
quantifying the contribution of $K^*$-resonance in the $B\to K\pi \ell^+\ell^-$ decays.  
All axial-vector and tensor $B\to K\pi $ form factors can in principle 
be calculated, choosing an appropriate  $b\to s$ transition current 
in the vacuum $\rightarrow K\pi$ correlation function similar to 
Eq.~(\ref{eq:corr}).
Here however one needs additional studies of kaon-pion DAs,
taking into account the  $SU(3)_{flavour} $ violating asymmetry
in the  Gegenbauer expansion, and establishing the accurate inputs for the 
coefficients which will involve various timelike  $ K\pi$ form factors. 


Further improvements of LCSRs obtained in this paper are possible 
in several directions: (1)  working out and taking into account 
the higher-twist components for the vacuum $\to$ dipion bilocal
matrix elements, most importantly the twist-3 DAs;
(2) calculating the gluon radiative corrections to the hard-scattering
amplitude and (3) performing a dispersion relation analysis for 
the coefficients of DAs considered as analytic functions 
of the dipion mass. 

Let us particularly discuss the future perspectives to
go beyond the twist-2 approximation in the LCSRs, such as Eqs.(23) and (24). 
To that end, one has to retain all operator structures
in the vacuum $\rightarrow$  dipion
matrix element (\ref{eq:corr3}) and identify 
their twist-3,4 components.
The latter have to be parametrized in terms of new DAs 
for which a double (conformal and spatial partial-wave) expansion has to be 
worked out, similar to Eq.~(\ref{eq:daexpansion}) used for the twist-2 DAs.
For the isospin-one dipion system, a systematic study of 
higher-twist effects should go along the similar
lines as in the analysis of $\rho$-meson DAs of 
twist-3,4 (see e.g., \cite{Ball:1998sk}), so that 
the role of  the polarization four-vector of the vector meson 
will be played by the difference of four-momenta 
$\bar{k}$. 
The emerging coefficients
of twist-3,4 DAs -- 
analogs of the Gegenbauer coefficients $B_{n\ell}^{\perp,\parallel}(k^2)$-- 
will represent 
new timelike pion form factors of certain local (twist-3,4) operators. 
Note that similar to the twist-2 coefficients, 
these will be complex functions at $k^2\geq 4m_\pi^2$. 
Hence, as opposed to the parameters of one-pion DAs,
one cannot access the dipion DAs using QCD sum rules with local OPE.
The only timelike form factor available from experiment 
is the pion electromagnetic form factor $F_\pi(k^2)$ 
determining the coefficient   
$B_{10}^{\parallel}(k^2)$.
To obtain the remaining coefficients $B_{n>1,\ell}^{\parallel}(k^2)$,
$B_{n,\ell}^{\perp}(k^2)$ of twist-2 DAs 
and the new emerging coefficients  of the twist-3,4 DAs
one has to combine theoretical methods with the data on two-pion 
scattering in different partial waves. 
Apart from the low-energy QCD calculations  such as 
the instanton model at low $k^2$ \cite{PolyakovWeiss} we used for 
the DA coefficients here, a promising strategy to access 
the larger $k^2\lesssim 1$ GeV$^2$ region is
to apply hadronic dispersion relations for the coefficients 
of DAs in the variable $k^2$, as suggested already in 
\cite{PolyakovNP}. These relations will involve known 
resonance structure (positions and widths of two-pion resonances) 
and can make use of pion scattering phases (via Omnes representation,
see e.g.,~\cite{PolyakovNP} and \cite{Kubisetal}), but need additional
input for normalization of the resonance residues and/or    
subtraction constants.
One possibility to fix the normalizations is to employ 
dedicated LCSRs with one-pion DAs and the pion interpolating current,
similar to the LCSRs for the pion electromagnetic 
form factor \cite{Braun:1999uj}.
These auxiliary sum rules will allow one to calculate the new form factors 
related to the coefficients of dipion DAs in the spacelike region of $k^2$. 
Afterwards, one  fits the parameters  
in the hadronic dispersion relations matching the latter   
at $k^2<0$ to the LCSR calculation. 
This kind of matching between LCSR results and dispersion representation
works for the pion electromagnetic form factor, as discussed in 
\cite{Bruch:2004py}. 
We plan a  dedicated study along these lines.


With the LO and twist-2 accuracy, the sum rules for $B\to 2\pi$ form factors 
obtained in this paper, represent the first exploratory step towards 
further development of the new LCSR method and towards its extensions to the other 
important hadronic heavy-to-light form factors with two mesons
in the final state.

\vspace{-0.3cm}
\section*{Acknowledgments} 
We are grateful to Danny van Dyk, Thorsten Feldmann, Thomas Mannel and Javier Virto 
for useful discussions. This work is supported by 
DFG Research Unit FOR 1873 
``Quark Flavour Physics and Effective Theories'',  Contract No.~KH 205/2-1. 

\begin{appendix}
\section*{Appendix A: Details on kinematics} 
\label{sect:app1}

The correlation function (\ref{eq:corr}) can formally be viewed as an amplitude of 
a $2\to 2$ process, in which the two initial particles 
with the squared masses $p^2$  and $q^2$ produce a final dipion state,  
The dipion mass squared  $k^2=(p-q)^2$ plays then the role of 
the Mandelstam variable $s$, so that $k^2\geq 4m_\pi^2$, 
whereas $ q\cdot \overline{k}= p\cdot \overline{k}=(t-u)/2$, and  
the standard condition for the sum of the 
three variables reads: $s+t+u=2m_\pi^2+q^2+p^2$.  
The following kinematical limits for the variable  $ q\cdot \overline{k}$ 
are then derived using a general inequality for the Mandelstam variables:
\be
-\lambda^{1/2}(p^2,q^2,k^2)\sqrt{1-\frac{4m_\pi^2}{k^2}}
\leq 2(p\cdot \overline{k})\leq \lambda^{1/2}(p^2,q^2,k^2)\sqrt{1-\frac{4m_\pi^2}{k^2}}\,,
\label{eq:cond}  
\ee
where $\lambda(a,b,c)=a^2+b^2+c^2-2ab-2ac-2bc$.

It is convenient to decompose the momenta $k$, $k_{1,2}$ near the light cone:
\be  
k^\mu=\frac12(k^+n^{+\mu}+k^-n^{-\mu})+k^{\perp \mu}\,,~~
\label{eq:LCdecomp}
\ee
where 
$n^{\pm \mu}=(1,0,0,\pm 1)$. 

The  parameter 
$\zeta$ determines the light-cone momentum fractions carried  
by the two pions in the final state \cite{2pionDA,PolyakovNP}:
\be
\zeta=k_1^+/k^+,~~ 1\!-\!\zeta=k_2^+/k^+\,,~~\zeta(1-\zeta)\geq \frac{m_\pi^2}{k^2}\,.  
\label{eq:zeta}
\ee
To relate this parameter to the invariant variable \,$q\cdot\overline{k}$,
it is convenient to choose the kinematical configuration where 
the four-momenta $p$ and $q$ of external currents in the correlation function 
are aligned with the $z$-direction, so that $k^{\perp \mu}=0$.
The relation has then a form of quadratic equation
with a solution:
\be
q\cdot \bar{k}=\frac12(2\zeta -1)\lambda^{1/2}(p^2,q^2,k^2)\,,
\label{eq:qkbar} 
\ee
At $p^2=m_B^2$
we recover the relation (\ref{eq:qkbar}) for $B\to \pi\pi
\ell\nu_\ell $ decay (see e.g., \cite{FFKMvD}). 
The parameter $\zeta$ is related via  Eq.~(\ref{eq:zetaTheta})
to the angle  between the pions  in their c.m. frame.
The latter
relation substituted in   Eq.~(\ref{eq:qkbar}) reproduces 
the limits (\ref{eq:cond}). 

The origin of the imaginary part of the $\lambda^{1/2}$-function in the variable $p^2$ 
mentioned in Sect.2  is evident from the following form:
\be 
\lambda^{1/2}(p^2,q^2,k^2)= (p^2-(\sqrt{q^2}-\sqrt{k^2})^2)^{1/2}(p^2-(\sqrt{q^2}+\sqrt{k^2})^2)^{1/2}.
\ee

\section*{Appendix B: Details on dipion DA's} 
\label{sect:app2}
The coefficient functions of the double polynomial expansion of dipion DAs 
are multiplicatively renormalized in the one-loop approximation: 
\begin{align}\label{eq:renorm}
B_{nl}^{\parallel,\perp}(k^2,\mu)=B_{nl}^{\parallel,\perp}(k^2,\mu_0)
\Big(\frac{\alpha_s(\mu)}{\alpha_s(\mu_0)} \Big)^{(\gamma_n^{\parallel,\perp}-\gamma_0^{\parallel,\perp})/\beta_0}\,,
\end{align}
where $\beta_0=11-2/3n_f$, 
and the anomalous dimensions are \cite{renorm}:
\begin{align}\label{eq:anomdim}
\gamma_n^{\parallel}= C_F\Big(1-\frac{2}{(n+1)(n+2)}+4\sum\limits_{k=2}^{n+1}
\frac{1}{k}  \Big), ~~
\gamma_n^{\perp}=\frac{8}{3}\Big(1+4\sum\limits_{k=2}^{n+1}\frac{1}{k}  \Big)\,.
\end{align} 

For the chiral-odd dipion DA these functions are taken from \cite{PolyakovNP} 
where they are calculated at small $k^2$ in the instanton model of QCD vacuum
at the scale $\mu\simeq 600$ MeV:
\ba
B_{01}^\perp(k^2)&=&1+\frac{k^2}{12 M_{0}^2},
\nonumber\\
B_{21}^\perp(k^2)&=&\frac{7}{36}\left(1-\frac{k^2}{30 M_{0}^2}   \right),~~
B_{23}^\perp(k^2)=\frac{7}{36}\left(1+\frac{k^2}{30 M_{0}^2}   \right),
\label{eq:Bnl}\\
B_{41}^\perp(k^2)&=&\frac{11}{225}\left(1-\frac{5 k^2}{168 M_{0}^2}   \right),~
B_{43}^\perp(k^2)=\frac{77}{675}\left(1-\frac{k^2}{630 M_{0}^2}   \right),~
B_{45}^\perp(k^2)=\frac{11}{135}\left(1+\frac{k^2}{56 M_{0}^2}   \right)\,.
\nonumber
\ea
The normalization constant is  related to the 
key mass parameter of the instanton model $M_{0}\simeq 350$ MeV via 
$f^\perp_{2\pi}=4\pi^2f_\pi^2/3M_0\simeq 650$ MeV, where 
$f_\pi=132 $ MeV is the pion decay constant. 
\end{appendix}

\end{document}